\documentclass[%
 reprint,
]{revtex4-2}

\usepackage{graphicx}
\usepackage{caption,subcaption}
\usepackage{dcolumn}
\usepackage{bm}

\usepackage{amsthm,amsmath,amsfonts,amssymb}
\usepackage{mathtools,mathrsfs}
\usepackage{setspace}
\usepackage{xcolor}
\usepackage{thmtools} 

\usepackage[pdfdisplaydoctitle,colorlinks,breaklinks,urlcolor=blue,linkcolor=blue,citecolor=blue]{hyperref} 
\usepackage[nameinlink]{cleveref}

\newcommand{\R}{\mathbb{R}}

\newcommand{\set}[1]{\left\{#1\right\}}
\newcommand{\pa}[1]{\left(#1\right)}
\newcommand{\bra}[1]{\left[#1\right]}
\newcommand{\abs}[1]{\left|#1\right|}

\newcommand{\brak}[1]{\left\langle#1\right\rangle}

\newtheorem{theorem}{Theorem}[section]

\newtheorem{proposition}[theorem]{Proposition}

\theoremstyle{remark}
\newtheorem{remark}[theorem]{Remark}
\numberwithin{equation}{section}

\begin{document}

\preprint{APS/123-QED}

\title[Zero-Noise for Point Vortex Dynamics after Collapse]{Zero-Noise Selection 
		for Point Vortex Dynamics after Collapse}

\author{Francesco Grotto}
 \email{francesco.grotto at unipi.it}
\affiliation{%
 Dipartimento di Matematica, Universit\`a di Pisa\\ Largo Pontecorvo 5, 56127 Pisa, Italia
}%

\author{Marco Romito}
\email{marco.romito at unipi.it}
\affiliation{%
 Dipartimento di Matematica, Universit\`a di Pisa\\ Largo Pontecorvo 5, 56127 Pisa, Italia
}%

\author{Milo Viviani}
\email{milo.viviani at sns.it}
\affiliation{
 Scuola Normale Superiore, Classe di Scienze\\ Piazza dei Cavalieri, 7, 56126 Pisa, Italia
}%

\date{\today}

\begin{abstract}
The continuation of point vortex dynamics after a vortex collapse is investigated by means of a regularization procedure consisting in introducing a small stochastic diffusive term, that corresponds to a vanishing viscosity. 
In contrast with deterministic regularization, in which a cutoff interaction selects in the limit a single trajectory of the system after collapse, the zero-noise method produces a probability distribution supported by trajectories satisfying relevant conservation laws of the point vortex system.
\end{abstract}


\maketitle

\section{Introduction}

The point vortex (PV) system,
\begin{equation}\label{eq:pv}\tag{PV}
    \dot x_i=\frac1{2\pi} \sum_{j\neq i}^N \gamma_j \frac{(x_i-x_j)^\perp}{\abs{x_i-x_j}^2},
\end{equation}
describing the evolution of vortices with positions $x_1,\dots,x_N\in\R^2$ and intensities $\gamma_1,\dots,\gamma_N\in\R\setminus\set{0}$,
is a classical model dating back to the works of Helmholtz \cite{Helmholtz1858}, describing the dynamics of an idealized 2-dimensional incompressible fluid for which vorticity is concentrated in a finite number of points. It is a Hamiltonian ODE system with singular interactions between particles, and it is a well-studied model for both its mathematical properties and physical interest as a discretization method in fluid dynamics \cite{Chorin1973}.

The vector field driving \eqref{eq:pv} is singular when the positions of two or more vortices coincide. Singular configurations of point vortices leading to collapse in finite time, or (equivalently, by time inversion) the evolution of many vortices starting from the same position (\emph{vortex burst}), have been known since the XIX century \cite{Aref2010}. A classical result by D\"urr-Marchioro-Pulvirenti \cite{Durr1982,Marchioro1994} states that these singular configurations are a negligible set of phase space with respect to absolutely continuous invariant measures of the system, that is there exists a measure-preserving flow on phase space consisting of smooth solutions  \cite{Grotto2020}. Nevertheless, there exists a large variety of singular configurations, which are in fact dense in phase space.

It was proved in \cite{Grotto2022a} that given any configuration of $N$ point vortices at time $t=0$, there exists a solution of \eqref{eq:pv} with $N+2$ vortices, with three of them bursting out of a single one at time $t>0$. In the same way, it is possible to construct configurations of arbitrarily many vortices leading to collapse in finite time, and this kind of singular configurations contains codimension-1 submanifolds of phase space. This reveals the existence of a rich variety of vortex collapses and bursts, opening the possibility of PV models for vorticity creation and energy dissipation in incompressible fluids \cite{Leoncini2011}.
Moreover, advection by the velocity field induced by PV systems close to collapse can be regarded as a model for the chaotic Lagrangian movement of particles in turbulent 2D flows \cite{Leoncini2000,Leoncini2001}, providing a further motivation for the study of singular vortex configurations.

Considering PV dynamics as a generalized solution to 2D incompressible Euler equations (\emph{cf.} \Cref{ssec:weaksol} below), these facts imply a wild non-uniqueness of the time evolution: for instance the construction in \cite{Grotto2022a} can be used to exhibit PV dynamics in which vortices split at completely arbitrary times.
Therefore, the question of what should be the correct (physical) continuation of the evolution after a vortex collapse, or before a vortex burst, naturally arises.

In this paper, we investigate the dynamics of $N=3$ vortices after their collapse through regularization by noise: we will introduce stochastic forcing terms in \eqref{eq:pv} that prevent collision by means of small-scale fluctuations, and then consider a \emph{zero-noise} limit, that is the (nontrivial) vortex dynamics obtained in the limit where the diffusion coefficient vanishes.
In particular, we will exhibit examples in which the zero-noise limit, after a collapse, is supported by many trajectories preserving first integrals of motion. In other terms, we will produce a PV evolution exhibiting a collapse at finite time, after which a \emph{spontaneously stochastic} (in the sense of \cite{Mailybaev2016,Thalabard2020}) continuation of the dynamics consists in a random choice of a bursting solution.

Our result will be compared to the (deterministic) regularization procedure in which the interaction kernel of vortices is modified so to prevent collapses \cite{Sakajo2012,Gotoda2016a,Gotoda2016b,Gotoda2018}, in which case a single possible trajectory is identified for the vortices to follow after collapse. More generally, such a procedure is a relevant tool in the investigation of ODE models after formation of singularities \cite{Drivas2021}.

The stochastic PV model we will consider is related (in the large $N$ Mean Field limit) to 2D Navier Stokes equations, for which our procedure corresponds to considering the inviscid limit. Therefore, the coexistence of the zero-noise selection principle we propose with other regularization procedures leading to different continuations for \eqref{eq:pv} after collapse indicates an intrinsic non-uniqueness of PVs as a 2D fluid dynamics model, whose relation with 2D turbulence remains open for future investigation.

\section{Vortex Collapse and Burst}

The Hamiltonian system of $N=3$ vortices on the whole plane $\R^2$ is integrable (\emph{cf.} \cite{Modin2021} for a general discussion on integrability of PV systems), and it admits singular solutions in which the vortices collapse in --or burst out of-- a single point, with self-similar trajectories, that is the angles of the triangle having as vertices the vortices' position are constants of motion.

\begin{theorem}\cite[Section 3]{Krishnamurthy2018}\label{thm:krishnamurthy}
	Consider the system \eqref{eq:pv}
	of $N=3$ vortices with intensities $\gamma_1,\gamma_2,\gamma_3\in\R\setminus\{0\}$ 
	starting from distinct positions $x_1,x_2,x_3\in\R^2$ at $t=0$.
	The existence time interval of a maximal solution of \eqref{eq:pv} ends with a collapse of the three vortices, or (alternatively) it begins with a burst, if and only if
	\begin{equation}\label{eq:condcollapse}
	\sum_{j \neq k}\gamma_j\gamma_k=0,\quad I=\sum_{j \neq k}\gamma_j\gamma_k |x_j-x_k|^2=0,
	\end{equation}
	and $x_1,x_2,x_3$ are not the vertices of an equilateral triangle.
	Conditions \eqref{eq:condcollapse} also imply that the motion of vortices is self-similar, and if starting positions form an equilateral triangle, the configuration is of relative equilibrium.
\end{theorem}

Let us remark that, while self-similarly collapsing configurations exist for more than three vortices \cite{Kudela2014}, for PV systems with $N\geq 4$ there also exist configuration leading to collapse with trajectories that are not self-similar \cite{Sakajo2008}.

The \emph{moment of inertia} $I(t)=I$ is a first integral of motion.
The first condition in \eqref{eq:condcollapse}
excludes that collapsing vortices, and thus bursts, have intensities of the same sign.
Indeed, in the latter case 	the Hamiltonian of the system,
\begin{equation*}
H=-\frac{1}{2\pi} \sum_{j\neq k} \gamma_j \gamma_k \log |x_j-x_k|,
\end{equation*}
(with conjugate coordinates $(x_{i,1},\gamma_i x_{i,2})_{i=1,\dots,N}$)
can be used to control the minimum distance of vortices.

Invariance of $H$ implies that there are no solutions in which two vortices collide at finite time and a third one remains distinct. The above Proposition thus characterises \emph{all} singular configurations for $N=3$, and it furthermore reveals that in those cases the motion is self-similar.

In particular, \emph{any continuation of vortex dynamics after collapse should be a self-similar burst}.
We are thus interested in characterizing the possible collapsing and bursting vortex configurations so to study how regularization procedures select among them a continuation of the dynamics.

\begin{figure*}
  \setlength{\unitlength}{\textwidth}
  \begin{picture}(1,0.185)
    \put(-0,-0.2){\includegraphics[width=11cm]{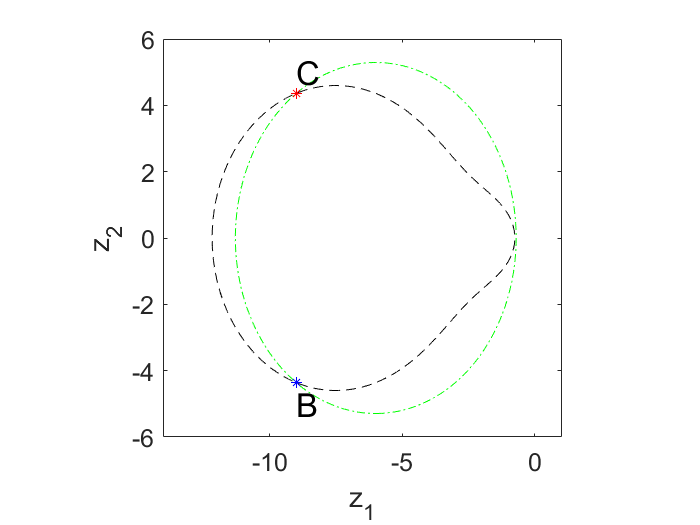}}
    \put(0.5,-0.18){\includegraphics[width=11cm]{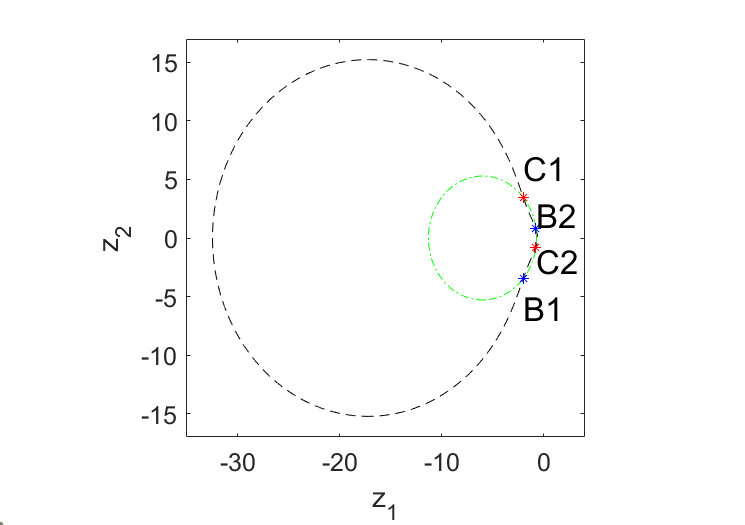}}
    \put(.657,-.1){\includegraphics[width=3.5cm]{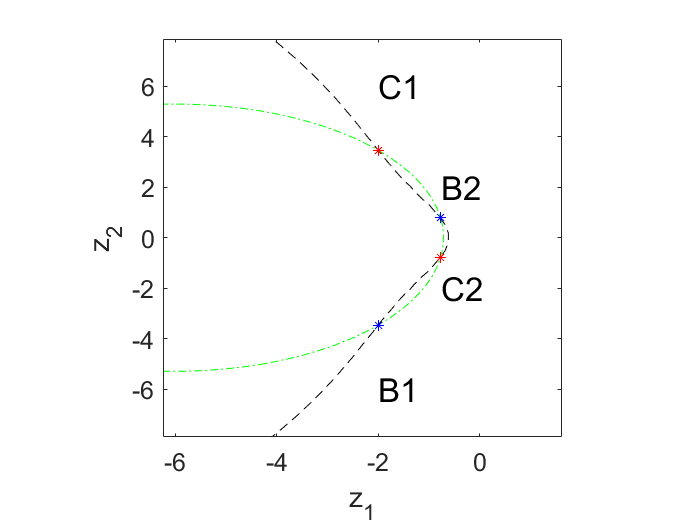}}
  \end{picture}
  \vspace{3cm}
\caption{\label{fig:energymomentum4} Intersections between the level sets $H=H_0$ (dashed black) and $I=0$ (solid green) for PV configurations with different intensities. On the left a case with only two intersections: the point C indicates an intersection corresponding to a collapsing configurations, and point B to a bursting configuration. On the left a case with four intersections, red and blue points marking respectively collapsing and bursting configurations.}
\end{figure*}

\subsection{Classifying Singular Configurations}\label{ssec:classifying}

Let us consider intensities $\gamma_1,\gamma_2,\gamma_3$ and initial positions $x_1(0),x_2(0),x_3(0)\in \R^2$ satisfying \eqref{eq:condcollapse}.
The center of vorticity $C=\gamma_1x_1+\gamma_2x_2+\gamma_3x_3$ is a first integral of PV motion, and we can assume that it is the origin $(0,0)$.
The self-similar solution then takes the form
\begin{multline}\label{eq:selfsimilar}
    x_i(t)=\sqrt{a (t^\ast-t)} 
    \Theta_{a,b}(t^\ast-t) x_i(0),\\
    \Theta_{a,b}(t)=
    \begin{pmatrix}
        \cos(\tfrac{b}{2a} \log(t)) & -\sin(\tfrac{b}{2a} \log(t))\\
        \sin(\tfrac{b}{2a} \log(t)) & \cos(\tfrac{b}{2a} \log(t))\\
    \end{pmatrix},
\end{multline}
where $a>0,b\neq 0$ are dynamical parameters that are determined completely by vortex intensities and initial positions (\emph{cf.} \cite[Section 2]{Grotto2022a}).
While we expect that the number and intensities of vortices be preserved in the continuation of dynamics after collapse, as we understand the latter as a limit of a regularized system for which that is the case,
in principle parameters ruling the self-similar evolution might be influenced by the collapse, namely the triangle of vortex positions could change its symmetry class. 

Exploiting the integrability of the system, we reduce the study of those parameters to that of a single point of $\R^2$.
Rescaling intensities corresponds to scaling time in \eqref{eq:pv}, so we can assume $\gamma_2=1$ without loss of generality, hence $\gamma_3=-\frac{\gamma_1}{1+\gamma_1}$
by the first condition of \eqref{eq:condcollapse}.
The collapse condition $I=0$ and the assumption $C=(0,0)$ of above give: 
\begin{eqnarray*} 
         & I=\gamma_1|x_1-x_2|^2-\frac{\gamma_1^2 |x_1-x_3|^2}{1+\gamma_1}-\frac{\gamma_1 |x_2-x_3|^2}{1+\gamma_1}=0\\
         & C=\gamma_1x_1 + x_2 -\frac{\gamma_1}{1+\gamma_1}x_3= (0,0).
\end{eqnarray*}
Excluding the trivial solution $x_1=x_2=x_3=0$, 
we can exploit invariance under rotations and homotheties of those conditions by applying $R\in  O^+(2)$ (an orthogonal matrix with positive determinant) such that $R x_3=(1,0)$.
The second equation now gives
\begin{equation*}
    R x_2 = \pa{\frac{\gamma_1}{1+\gamma_1},0} - \gamma_1R x_1,
\end{equation*}
so we find the possible configurations in terms of $Z=R x_1$ by intersecting the two conditions $I=I(Z)=0$ and 
\begin{multline*}
    H_0=H(Z)=\gamma_1\log|Z-R x_2|\\-\frac{\gamma_1^2}{1+\gamma_1}\log|Z-(1,0)|-\frac{\gamma_1}{1+\gamma_1}\log|Rx_2-(1,0)|
\end{multline*}
for any fixed choice of $H_0$ and $\gamma_1$. Such intersection of curves only consists of isolated points $Z$
encoding 1) the similarity class of the triangle of vertices $x_1,x_2,x_3$, 2) the angle of rotation of such triangle around $(0,0)$,
3) the order with which $x_1,x_2,x_3$ appear as vertices of the triangle.

Notice in particular that, given $\gamma_1,H_0$, if $Z=(z_1,z_2)$ is an intersection so is $\tilde Z=(z_1,-z_2)$, 
the two points corresponding to similar triangles with vertices $x_1,x_2,x_3$ in a different order, one leading to a vortex collapse and the other to a burst.

\emph{The self-similar evolution of vortices away from the burst/collapse time leaves $Z=Z(t)$ invariant}. In other words, this procedure classifies the types of self-similar evolution leading to a burst or collapse of vortices.
\Cref{fig:energymomentum4} reports two different cases, in which $H,\gamma_1$ are chosen so that there are 2 or 4 intersections.

\begin{remark}
    As suggested by the comparison between the plots in \cref{fig:energymomentum4}, it is possible that curves $H=H_0$ and $I=0$ in the $Z$-plane intersect in three points, as a limit case in which $B_2$ and $C_2$ coincide. In that case, the third point corresponds to a relative equilibrium, in which the PV configuration rotates periodically around its center of vorticity. Such dynamics, when perturbed by additive noise, persists in the zero-noise limit.
\end{remark}

We refer to \cite{Demina2014,Kallyadan2022} concerning procedures for determining singular PV configurations.

\subsection{Deterministic Regularization}

A natural way of investigating continuation after a vortex collapse is to regularize the vortex interaction, so that the modified ODE system is well-posed for all times, and then study the limit behavior in the regularization parameter. Consider a convolution kernel $h:\R\to\R$ and, for $\epsilon>0$, the modified PV dynamics
\begin{gather}\label{eq:rpv}
    \dot x_i^\epsilon=\sum_{j\neq i}^N \gamma_j K_\epsilon(x_i^\epsilon-x_j^\epsilon),\\ \nonumber
    \quad K_\epsilon=\nabla^\perp\bra{h^\epsilon\ast \pa{-\tfrac1{2\pi}\log}},\quad 
    h^\epsilon(x)=\frac1{\epsilon^2}h\pa{\frac{x}\epsilon},
\end{gather}
with $K(x)=\frac{x^\perp}{2\pi |x|^2}$ being the original interaction kernel.
The most relevant examples of kernels $h$ are: bump functions (that is, smooth compactly supported functions); the Gaussian probability distribution, $\frac1{2\pi} e^{-|x|^2/2}$; the Cauchy probability distribution, $(1+|x|^2)^{-2}/\pi$ (in this case \cref{eq:rpv} is the \emph{blob-vortex system}); $\frac1{2\pi}K_0(|x|)$ with $K_0$ the modified Bessel function of second kind (in this case \eqref{eq:rpv} is the $\alpha$\emph{-Euler vortex system}, \emph{cf. \cite{Gotoda2016b}}).
More generally, one can consider the a kernel $h$ satisfying the assumptions in \cite[Theorems 1,6]{Gotoda2018}.

For $\epsilon>0$, any choice of intensities $\gamma_1,\dots, \gamma_N\in\R\setminus\set{0}$ and initial positions $x_1(0),\dots, x_N(0)\in\R^2$, the ODE system \eqref{eq:rpv} is globally well-posed, the solution flow is a group of homeomorphisms that preserve the product Lebesgue measure on $(\R^2)^N$, and for all times $t\in\R$ and $i\neq j$, $x_i(t)\neq x_j(t)$, that is there are no collapses of vortices, \cite[Proposition 3]{Gotoda2018}.

Let now $N=3$, and consider a configuration $x_1(0),x_2(0), x_3(0)\in\R^2$ that leads to a collapse of the system \eqref{eq:pv} at time $t^\ast>0$. The solution $x_1^\epsilon(t),x_2^\epsilon(t), x_3^\epsilon(t)$ starting from the same initial configuration converges, as $\epsilon\to 0$, to the one of \eqref{eq:pv} for $t\in [0,t^\ast]$. Moreover, for $t> t^\ast$, $x_1^\epsilon(t),x_2^\epsilon(t), x_3^\epsilon(t)$ converges to a solution of \eqref{eq:pv} in which vortices start from the same position (\emph{burst}) and satisfy for $t\in (t^\ast,2t^\ast)$
\begin{equation*}
        (x_{i,1}(t),x_{i,2}(t))=(x_{i,1}(2t^\ast-t),-x_{i,2}(2t^\ast-t)).
\end{equation*}
We refer to \cite[Theorems 1,9]{Gotoda2018} for the latter facts.
Indeed, at fixed $\epsilon>0$, vortices in the regularized system become collinear at time $t^\ast$, and vortex positions converge to the same point at that time as $\epsilon\to 0$. The result we just described thus selects a precise continuation of the vortex dynamics after collapse, characterized by the symmetry encoded in the last displayed formula.
In terms of the discussion of the previous paragraph, the selected solution is such that the triangle of vertices $(x_1,x_2,x_3)$ remains self-similar before and after the collapse time $t^\ast$, but after the collapse the triangle has changed its orientation, that is $Z=Z(t)$ is changed to $\tilde Z$ after collapse.
Referring to \cref{fig:energymomentum4}, this corresponds to moving from point $C$ to $B$ in the figure on the left, and from $C_1$ to $B_1$, respectively $C_2$ to $B_2$, in the figure on the right.

\section{Stochastic Regularization}

The introduction of a common multiplicative stochastic forcing prevents the collapse of vortices given any arbitrary initial configuration: the stochastic forcing in \cite{Flandoli2011} aimed at modelling 2D Euler dynamics driven by transport noise, motivated by the regularizing effect of the latter in different contexts, such as ODE models \cite{Delarue2014} and advection of scalars by irregular vector fields \cite{Flandoli2010}.

We will consider instead additive (2-dimensional) Brownian noise acting independently on single particles, that is we consider the stochastic PV system
\begin{equation}\label{eq:pvadditivenoise}\tag{SPV}
    \dot x_i^\epsilon=\sum_{j\neq i}^N \gamma_j K(x_i^\epsilon-x_j^\epsilon)+\epsilon \dot B^i,
    \quad \epsilon>0,
\end{equation}
that is we add to the dynamics of each particle an isotropic stochastic forcing that perturbs trajectories but does not introduce bias towards a single direction. 
This PV system would model 2D Navier-Stokes equations in the mean field limit as $N\to \infty$, \cite{Marchioro1982}. 
In fact, we will consider systems of only 3 vortices, and our choice of stochastic forcing is due to the fact that additive noise is a simpler mathematical model (\emph{i.e.} it introduces a single additional parameter) while retaining the regularization effect. In particular, it holds:

\begin{proposition}
    Assume that $B^1_t,\dots,B^N_t$ are independent standard Brownian motions defined on a fixed probability space, and that intensities $\gamma_i$ satisfy
    \begin{equation*}
        \text{for all } I \subseteq\{1, \ldots, N\}, \quad \sum_{i \in I} \gamma_i \neq 0 .
    \end{equation*}
    Let $\epsilon>0$; given any initial condition $x_1(0),\dots, x_N(0)$, the singular SDE system
    \eqref{eq:pvadditivenoise} has a pathwise unique, strong solution for $t\geq 0$.
\end{proposition}

The latter result was established \emph{for Lebesgue-almost all initial conditions} in \cite[Theorem 1.1]{Fontbona2007}, extending the result on stochastic PV systems of \cite[Section 3]{Durr1982} on the 2-dimensional torus to the case of vortices on the whole plane. The arguments of \cite[Section 3.6]{Flandoli2011}, relying on Markov property and Liouville invariance of Lebesgue measure on phase space, allow to obtain well-posedness of \eqref{eq:pvadditivenoise} for \emph{any} initial configuration, since the case of additive noise we consider is actually simpler than the multiplicative one of that work. 

We will be interested in the limit of the dynamics \eqref{eq:pvadditivenoise} in the limit $\epsilon\to 0$. It is not difficult to show that, before the collapse time $t^\ast$ of the deterministic system, \eqref{eq:pvadditivenoise} pathwise converges to the latter. However, the system may not converge pathwise after collapse time, that is for a fixed sample of the driving noise trajectories $x_i^\epsilon(t)$ may not have a limit as $\epsilon\to 0$. Numerical approximations methods described below exhibit wide oscillations of single trajectories in such limit. This is due to the irregularity of Brownian samples, which is nevertheless essential for regularization: more regular (say, Lipschitz) forcing does not prevent collapse (\emph{cf.} \cite{Grotto2022a}).

\subsection{Limiting Dynamics in the Zero-Noise Limit}\label{ssec:weaksol}

\begin{figure*}
    \begin{subfigure}{.5\textwidth}
    \includegraphics[width=\textwidth]{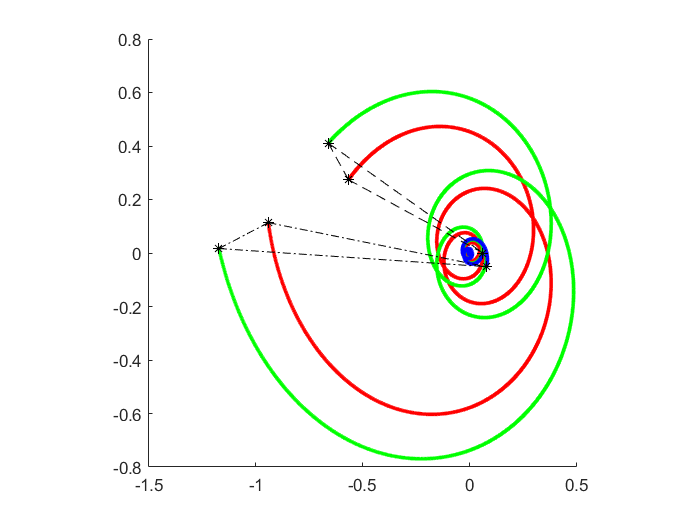}
    \end{subfigure}%
    \begin{subfigure}{.5\textwidth}
    \includegraphics[width=\textwidth]{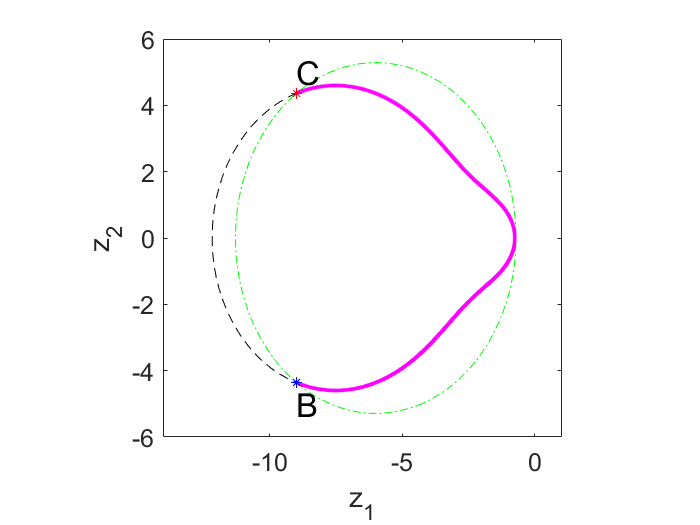}
    \end{subfigure}
    \caption{A single sample of a solution of \eqref{eq:pvadditivenoise} with $\epsilon=10^{-10}$. On the left, trajectories of vortices form a self-similar burst with the positions' triangle (dashed black) being similar to that of the initial configuration (point-dash black). On the right, the trajectory of $Z(t)$ (solid purple), closely following the energy level set from one intersection to another. For this initial configuration, there exist only one similarity class for the outgoing burst.    
    \label{fig:traj2}}
\end{figure*}

PV dynamics can be regarded as weak solutions of 2D Euler equations in vorticity form,
\begin{equation*}
    \partial_t \omega+ (K\ast\omega)\cdot \nabla \omega=0,
\end{equation*}
where $\omega=\nabla^\perp \cdot u$ is the vorticity corresponding to the velocity field of the fluid, which can be expressed by Biot-Savart law as $u=K\ast \omega$. Schochet \cite{Schochet1996} argued that PV systems satisfy Euler equations in the form:
\begin{align}
    \nonumber
    \omega(t)&=\sum \gamma_i \delta_{x_i(t)}, \quad \brak{\phi,\omega(t)}=\sum \gamma_i\phi(x_i),\\
    \label{eq:shochet}
    \partial_t \brak{\phi,\omega_t}&=\brak{H_\phi,\omega_t \otimes \omega_t},\\
    \nonumber
    H_\phi(x,y)&=\tfrac12 (\nabla\phi(x)-\nabla\phi(y)) K(x-y),
\end{align}
where $H_\phi$ is a bounded function (as revealed by Taylor expanding the observable $\phi$) which is discontinuous at $x=y$, where one postulates $H_\phi(x,x)=0$, and \eqref{eq:shochet} holds for any regular solution of Euler equations. As already remarked by Schochet, these measure-valued PV solutions are wildly non-unique because burst and collapses of vortices are admitted. It is worth mentioning, however, that in the context of Statistical Mechanics of PV ensembles (\emph{cf.} \cite{Geldhauser2019} for an overview) such weak notion of solution is so far the only means to study limiting Eulerian dynamics of PV as $N\to\infty$ \cite{Romito2020a,Romito2020b}.

One can prove (\emph{cf.} \cite[Proposition 4.5]{Grotto2022a}) that as long as vortices do not collide any such solution must correspond to empirical measure dynamics of a solution of \eqref{eq:pv}. However, say in the case of $N=3$ points collapsing at time $t^\ast$, not only any burst of vortices is admissible as a solution of \eqref{eq:shochet} as long as the total circulation is preserved, that is also possibly changing the number of vortices, but the trivial solution in which vortices remain stuck to the collapse position as a single still vortex is also \emph{a priori} a possibility\footnote{Rigorously speaking, we are admitting as solutions weak*-continuous curve of measures that satisfy \eqref{eq:shochet}.}. 

\emph{The zero-noise regularization procedure we propose produces, in the limit $\epsilon\to 0$ and after collapse time $t^\ast$, a probability distribution on solutions of \eqref{eq:pv} bursting out of the collapse point.}
Although this approximation selects more general continuations after collapse than the deterministic procedure outlined above, it provides strong evidence against vortices coalescing into a single one for $t>t^\ast$, that is vortices always burst out of the collapse position.

\begin{figure*}
\begin{subfigure}{.5\textwidth}
  \includegraphics[width=\textwidth]{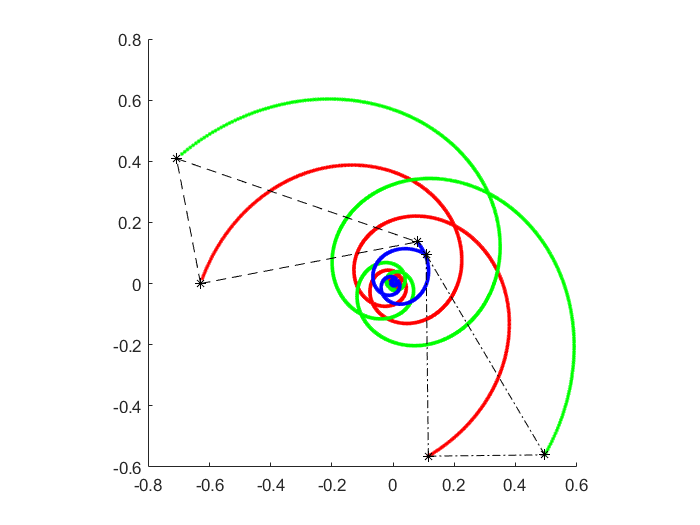}
\end{subfigure}%
\begin{subfigure}{.5\textwidth}
  \includegraphics[width=\textwidth]{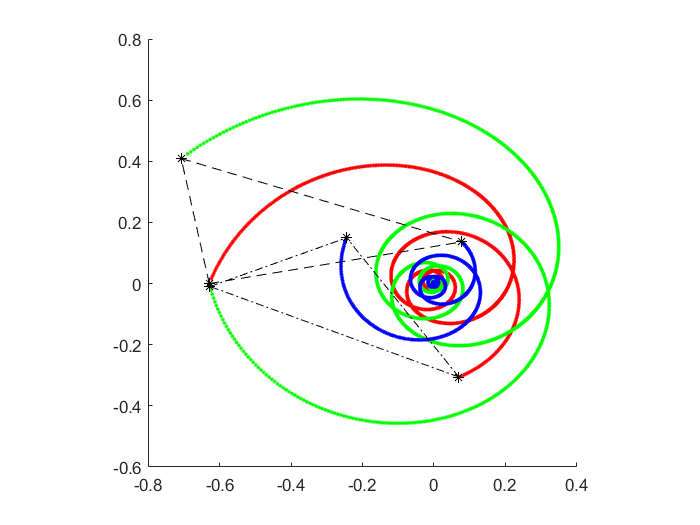}
\end{subfigure}
\begin{subfigure}{.5\textwidth}
  \includegraphics[width=\textwidth]{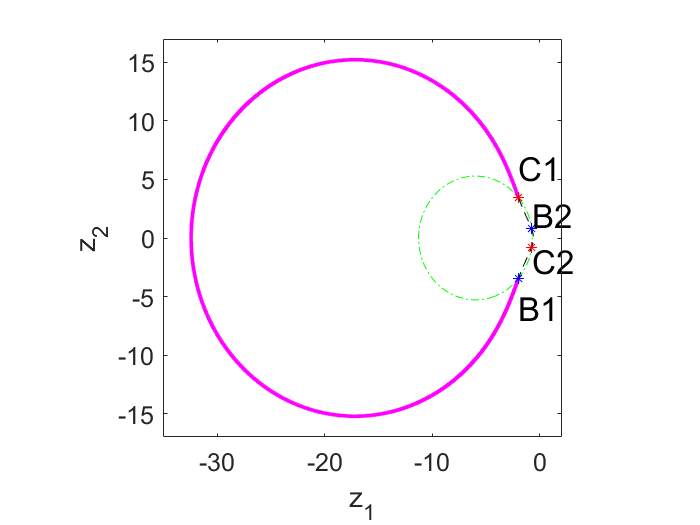}
\end{subfigure}%
\begin{subfigure}{.5\textwidth}
  \includegraphics[width=\textwidth]{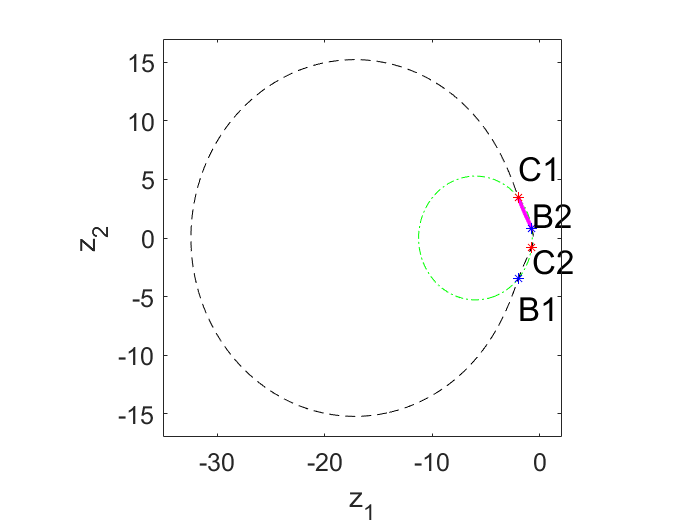}
\end{subfigure}
  \setlength{\unitlength}{\textwidth}
  \begin{picture}(1,0.185)
    \put(.618,.26){\includegraphics[width=3.2cm]{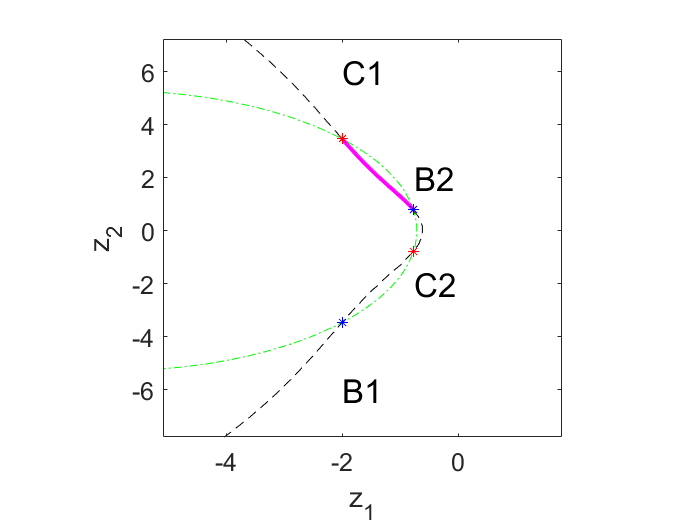}}
  \end{picture}
  \vspace{-3cm}
\caption{Two trajectories of \eqref{eq:pvadditivenoise} in the case where the self-similar burst after collapse can have two distinct similarity classes. Color legenda as in \cref{fig:traj2}. \label{fig:traj4}}
\end{figure*}

\subsection{Numerical Approximation of Stochastic Dynamics}

The main contribution of this paper is a quantitative numerical investigation of the zero-noise limiting behavior of \eqref{eq:pvadditivenoise}. Before moving to the main results on zero-noise statistics we briefly discuss numerical methods for approximating the system, that is integration schemes which are able to approximate with precision the vortex motion when the vector field becomes large due to vortices approaching each other.

A first possibility is to consider a SDE modified via the so called \emph{Darboux–Sundman transformation}. We define a new Hamiltonian
\begin{multline*}
    \tilde H(x_1,x_2,x_3) = \sigma(x_1,x_2,x_3)^\alpha(H(x_1,x_2,x_3)-H_0),\\
    \sigma(x_1,x_2,x_3)=\frac{1}{\frac{1}{\|x_1-x_2\|^2}+\frac{1}{\|x_2-x_3\|^2}+\frac{1}{\|x_1-x_3\|^2}},
\end{multline*}
for some $\alpha\geq1$.
Then, $\tilde H$ is $C^1$ function and so the modified SDEs, for $i=1,2,3$,
\begin{equation*}
    \dot x_i = J^{-1}\nabla_{x_i} \tilde H(x_1,x_2,x_3)+ \epsilon \dot W_i,
\end{equation*}
equivalently
\begin{multline*}
    \dot x_i =\sigma(x_1,x_2,x_3)^\alpha J^{-1}\nabla_{x_i} H(x_1,x_2,x_3)\\
         +(H(x_1,x_2,x_3)-H_0)J^{-1}\nabla_{x_i} \sigma(x_1,x_2,x_3)^\alpha+ \epsilon \dot W_i,
\end{multline*}
 have continuous (actually log-Lipschitz) drift.
We can then define the symplectic midpoint method
\begin{multline}
    \label{eq:midpoint_method}
 x^{n+1}_i = x^{n}_i\\
 + h J^{-1}\nabla_{x_i} \tilde H \pa{\frac{x^n_1 + x^{n+1}_1}{2},\frac{x^n_2 + x^{n+1}_2}{2},
 \frac{x^n_3 + x^{n+1}_3}{2}}\\
 + \varepsilon(W^{n+1}_i - W^n_i).
\end{multline}
This integration method provides a good numerical approximation while retaining important properties such as small fluctuations of first integrals of motion. However, it does so by slowing the flow when vortices are near, so in order to observe the solution beyond the point of collapse a large number of time iterations becomes necessary. 
Furthermore, the Darboux–Sundman transformation modifies the original Hamiltonian in a way that the collapse of the vortices never occurs, even without noise. 
This makes this integration technique not suitable for our purposes, even though for $\varepsilon>0$ is possible to re-scale the time via $\sigma$ to extrapolate the desired statistics.

An alternative to the Darboux–Sundman transformation is to use the function $\sigma^\alpha$ to adapt the time-step $h$, reducing it when vortices are close to each other and thus leading to an integration scheme in which each time-step has a comparable contribution to vortex displacement.
In our experiments, we have set $\alpha=1.3$.
We thus use the midpoint method \eqref{eq:midpoint_method} with respect to $H$, and time-step $h:=h(n)$, such that $h(n)\sigma(x^n_1,x^n_2,x^n_3)^\alpha=h(0)\sigma(x^0_1,x^0_2,x^0_3)^\alpha$. 
Notice that this method still requires a large number of time iterations to observe vortices well past the collapse time, as the time-step drastically decreases in proximity of the latter.

The forthcoming results have been obtained through the adaptive method because it offers the convenience of not requiring a time rescaling, as in the first method described.

We also use the modified midpoint method, introduced in \cite{Burrage2014}, where two independent samples of the Wiener process are required per time-step:
\begin{equation}\label{eq:midpoint_method_2noise}
 \begin{array}{ll}
      K_i =& x^{n}_i
 + \frac{h}{2} J^{-1}\nabla_{x_i} \tilde H \pa{K_1,K_2,K_3}\\
& + \varepsilon(W^{1,n+1}_i - W^{1,n}_i)\\
      x^{n+1}_i =& x^{n}_i
 + h J^{-1}\nabla_{x_i} \tilde H \pa{K_1,K_2,K_3}\\
& + \varepsilon\frac{1}{\sqrt{2}}\pa{(W^{1,n+1}_i - W^{1,n}_i)+(W^{2,n+1}_i - W^{2,n}_i)}.
\end{array}   
\end{equation}
Unlike the standard midpoint method \ref{eq:midpoint_method}, the \ref{eq:midpoint_method_2noise}  has the advantage of producing the correct drift for quadratic separable Hamiltonian.

\subsection{Qualitative Properties of Solutions with Small Noise}

We will consider solutions of \eqref{eq:pvadditivenoise} starting from configurations that would lead to self-similar collapse for \eqref{eq:pv}. Consider such a configuration $x_1,x_2,x_3$ with intensities $\gamma_1,\gamma_2,\gamma_3$, and let $H_0=H(x_1,x_2,x_3,\gamma_1,\gamma_2,\gamma_3)$ be the associated energy level. The discussion of \cref{ssec:classifying} reveals there may exist more than one burst of three point vortices with intensities $\gamma_1,\gamma_2,\gamma_3$ such that their Hamiltonian equals $H_0$.

The numerical evidence we detail in the next Section suggests that 
\emph{the zero noise limit of} \eqref{eq:pvadditivenoise} \emph{after the collapse time $t^\ast$ consists of a probability measure supported by bursting self-similar solutions of} \eqref{eq:pv} \emph{such that their Hamiltonian equals} $H_0$. Condition $I=0$ also holds for the limiting process after collapse. In fact, both the Hamiltonian and $I$ are local martingales under the dynamics \eqref{eq:pvadditivenoise}, their stochastic differential being given by 
\begin{gather*}
    d H(x_1,x_2,x_3)= \sum_{i=1,2,3} \nabla_i H(x_1,x_2,x_3)\cdot \epsilon dB^i,\\
    d I(x_1,x_2,x_3)= \sum_{i=1,2,3} \nabla_i I(x_1,x_2,x_3)\cdot \epsilon dB^i.
\end{gather*}
Moment of inertia $I$ is a smooth observable of trajectories, so its conservation in the zero-noise limit of the stochastic system \eqref{eq:pvadditivenoise} is perhaps more expected than that of the energy $H$.

The quantity $Z(t)$ introduced in \cref{ssec:classifying} is preserved by the stochastic evolution \eqref{eq:pvadditivenoise}, and in the limit $\epsilon\to 0$ it converges to a piecewise constant function: since the motion in the zero-noise limit is self-similar both before and after collapse, $Z(t)$ equals its initial value until collapse time $t^\ast$, then it jumps to a different intersection of level sets of $I$ and $H$ corresponding to a bursting solution.

\Cref{fig:traj2,fig:traj4} show single samples of \eqref{eq:pvadditivenoise} for small $\epsilon=10^{-10}$. The initial Hamiltonian $H_0$ is such that there are respectively two or four intersection points of $H=H_0$ and $I=0$ (as in \cref{fig:energymomentum4}), so there are two similarity classes for the bursting solution after collapse. The two trajectories in \cref{fig:traj4} portray the two distinct similarity cases.

\Cref{fig:traj2,fig:traj4} also depict the evolution of $Z(t)$. As $\epsilon$ decreases, $Z(t)$ will remain closer to intersection points, performing its jump from a collapse point to a burst point in a shorter and shorter time interval around $t^\ast$. However, for $\epsilon>0$ we observe that $Z(t)$ moves from an intersection corresponding to a collapsing configuration to one associated with a burst closely (but not exactly) following the Hamiltonian level set. Conservation of moment of inertia is only fully recovered in the limit, indicating that conservation of the Hamiltonian is somewhat more robust to noise perturbation. Notice that $Z(t)$ follows different trajectories in order to reach different burst intersections in \cref{fig:traj4}.

In order to give a full description of the zero-noise limit of \eqref{eq:pvadditivenoise}
we are thus led to two quantitative questions: given an initial configuration $x_1,x_2,x_3$ of vortices with intensities $\xi_1,\xi_2,\xi_3$ that would lead \eqref{eq:pv} to a collapse at finite time $t^\ast$,
\begin{enumerate}
    \item if more than one similarity class is possible for the self-similar burst continuation after collapse, what is the limit probability of each class?
    \item Conditionally to the similarity class of the self-similar burst after time $t^\ast$, the zero-noise limit is thus supported by the rotation of a bursting solution of \eqref{eq:pv} by a random angle: what is the probability distribution of such angle?
\end{enumerate}
It is worth noticing that if $x_1^\epsilon(t),x_2^\epsilon(t),x_3^\epsilon(t)$ is a solution of \eqref{eq:pvadditivenoise}, the probability distribution of another solution starting from a rotated initial datum $Rx_1(0),Rx_2(0),Rx_3(0)$, $R\in SO(2)$, when observed at time $t$, will correspond to that of the former one rotated by $R$. This is due to the fact that the driving vector field commutes with rotations (because of its specific form $K(x)=\frac{x^\perp}{2\pi |x|^2}$) and invariance under rotation of the law of Brownian motion. 

We also observe that the collapse time $t^\ast$ has no particular meaning for a sample solution of \eqref{eq:pvadditivenoise}, but as $\epsilon$ gets smaller the time at which the three vortices are closest quickly approaches $t^\ast$. 

The next Section is devoted to the numerical study of the two outlined questions. 

\subsection{Concerning different regularization procedures}
The mathematical model under consideration lends itself to a number of natural generalizations, which we only briefly discuss, because of the many different possibilities, the difficulty in the analysis of some of them and especially the issue of their physical relevance.

A first, simple modification of the additive noise we added to the PV system consists in considering non-isotropic forcing, that is Brownian motions in \eqref{eq:pvadditivenoise} with covariance matrix differing from the identity. This might lead to a different distribution of the rotation angle of the burst after collapse. More generally, one might consider Brownian motions whose covariance differs from vortex to vortex, however this looks unreasonable unless the PV system is considered as part of a more complex fluid dynamical model.

A further option is to consider more complex stochastic forcing, such as the multiplicative noise in \cite{Flandoli2010},
\begin{equation*}
    \dot x_i=\sum_{j \neq i} \gamma_j K(x_i-x_j) +\sum_{m=1}^M \sigma_m (x_i) \circ \dot B_t^m
\end{equation*}
($\circ$ denoting Stratonovich stochastic integration) which can be regarded as a discretization of stochastically forced Euler equations.
In this case, the choice of vector fields $\sigma_m$ must both be physically meaningful and also comply with mathematical conditions discussed in \cite{Flandoli2010} for the dynamics to be well-posed even for singular initial configurations. In fact, to the best of our knowledge no such choice is explicitly available at present.

More generally, other random modifications of the PV system, consistent with fluid-dynamical PDE models in the macroscopic limit, might also prevent collapse and thus provide different regularization methods: we mention for instance random generation of new vortices \cite{Grotto2020b} or random choice of intensities \cite{GrottoPeccati}. 

\section{Zero-Noise Statistics}

Numerical simulations of \eqref{eq:pvadditivenoise} were performed by means of the integration method \eqref{eq:midpoint_method_2noise} with adaptive time-step as described above.
We considered two initial configurations (we report approximated positions):
\begin{align*}
    \label{eq:firstconf}\tag{A}
    x_1=(-9,\, -10.5),\quad x_2=(1,\, 4.36),\quad x_3=(6.54,\, 0)\\
    \label{eq:secondconf}\tag{B}
    x_1=(-9,\, 1),\quad x_2=(-10.5,\, 4.36),\quad x_3=(6.54,\, 0)
\end{align*}
both with the same choice of intensities,
\begin{equation*}
    \gamma_1=-1/2,\quad \gamma_2=1/3,\quad \gamma_3=-1.
\end{equation*}
Configurations \eqref{eq:firstconf} and \eqref{eq:secondconf} lead to a collapse for \eqref{eq:pv}, but the intersection between the energy level set $H=H_0$, $H_0$ being the Hamiltonian evaluated at the initial configuration, and $I=0$ consists in two points for \eqref{eq:firstconf} and of four points in \eqref{eq:secondconf}, just as depicted in \cref{fig:energymomentum4}.  

At $\epsilon=10^{-10}$, $10^4$ trajectory samples were run for initial data  \eqref{eq:firstconf} and $2\times 10^4$ for initial data \eqref{eq:secondconf}: this is because in the latter case a comparable number of samples for the two burst similarity classes was desirable. Values of $\epsilon$ smaller than $10^{-10}$ lead to similar results but are computationally more intensive and approach the machine epsilon, with numerical error potentially being a concern: extensive sampling was thus conducted at $\epsilon=10^{-10}$. Due to the adaptive (hence very small) time-step, higher order integration methods do not appear to improve the approximation.

\begin{figure}
    \includegraphics[width=.5\textwidth]{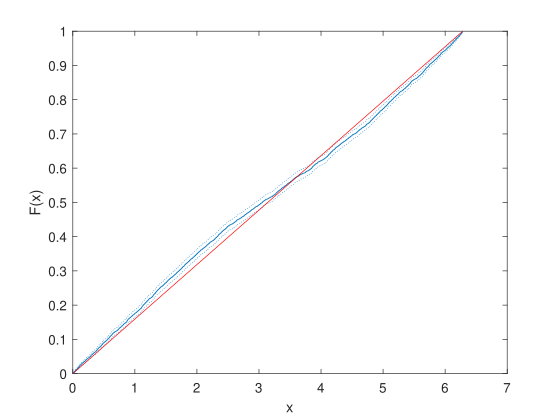}
    \caption{\label{fig:ecdf2} Empirical c.d.f. (solid blue) of the angle of $x_2^\epsilon(t^{\ast\ast})$ for $10^4$ samples of the dynamics \eqref{eq:pvadditivenoise} with $\epsilon=10^{-10}$ and initial configuration \eqref{eq:firstconf}. Dotted blue curves are confidence bounds ($95\%$) for the e.c.d.f. and the solid red line is the c.d.f. of the uniform distribution on $[0,2\pi]$.}
\end{figure}

\subsection{Equidistribution among symmetry classes}

With a sample size of $2\times 10^4$, trajectories of \eqref{eq:pvadditivenoise} after the deterministic collapse time $t^\ast$ were close to self-similar trajectories of two types, corresponding to the two burst intersections of energy and moment of inertia level sets, with 9926 trajectories in one class and 10074 in the other. Data is thus compatible with a uniform distribution between the two possibilities, with a $p$-value of 0.295. Simulations with different, larger values of $\epsilon$ (but reduced sample size) confirmed convergence to the uniform distribution.

\begin{figure*}
\centering
\begin{subfigure}{.5\textwidth}
  \centering
  \includegraphics[width=\textwidth]{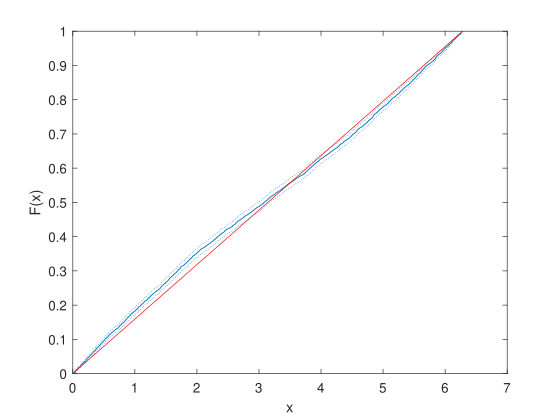}
\end{subfigure}%
\begin{subfigure}{.5\textwidth}
  \centering
  \includegraphics[width=\textwidth]{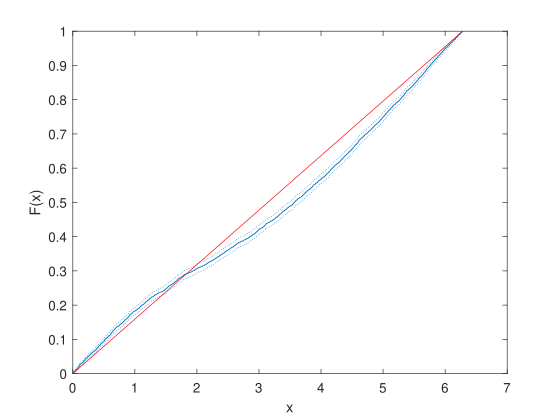}
\end{subfigure}
\caption{\label{fig:ecdf4} Empirical c.d.f. (solid blue) of the angle of $x_2^\epsilon(t^{\ast\ast})$ for $2\times 10^4$ samples of the dynamics \eqref{eq:pvadditivenoise} with $\epsilon=10^{-10}$ and initial configuration \eqref{eq:secondconf}, samples being divided (right and left plots) according to the similarity class of the PV position triangle during burst. Color legenda as in \cref{fig:ecdf2}.}
\end{figure*}

\subsection{Distribution of bursting configuration in each symmetry class}

In order to study the rotation angle of the self-similar bursting PV evolution after collapse in the zero-noise limit we considered the angle of vortex $x_2^\epsilon(t^{\ast\ast})$ (with respect to the origin, \emph{i.e.} the center of vorticity up to an error of order $10^{-9}$) measured at $t^{\ast\ast}=2t^\ast$, that is after the same time passed since the deterministic collapse time.

Measurement of the angle of $x_2^\epsilon$ was also considered at the (random) stopping time at which $|x_2^\epsilon|$ overcomes a fixed threshold, leading to similar results. This however requires a spline interpolation of trajectories in order to observe the position at the stopping time, possibly increasing numerical error, so we only report measurement at a given time. Another possibility consists in a parametric fit of \eqref{eq:selfsimilar} for determining the angle parameter $b/2a$ from many points in the trajectory of $x_2^\epsilon$ well after the deterministic collapse time; this produces again statistics similar to the fixed time observation while being more involved and potentially accumulating numerical error.

\Cref{fig:ecdf2} reports the empirical c.d.f. of the angle of $x_2^\epsilon(t^{\ast\ast})$ for initial configuration \eqref{eq:firstconf}, exhibiting a significant deviation from the uniform distribution. \Cref{fig:ecdf4} reports separately the c.d.f.'s of the same angle for initial configuration \eqref{eq:firstconf}, according to the similarity class of the positions' triangle at $t^{\ast\ast}$.

The same measurement was carried out in both cases \eqref{eq:firstconf},\eqref{eq:secondconf} for different values of $\epsilon$, both larger and smaller. However, the resulting empirical c.d.f.'s exhibit rather strong fluctuations as $\epsilon\to 0$, and it was not possible to extrapolate a limit distribution. Such behavior might be compatible with a non unique limit probability distribution. Nevertheless, the relatively small confidence bands reported in \cref{fig:ecdf2,fig:ecdf4} provide statistical evidence against uniform distribution, confirmed by negligible $p$-value ($\simeq 10^{-09}$, $\simeq 10^{-11}$ and $\simeq 10^{-46}$ respectively for empirical c.d.f.'s in \cref{fig:ecdf2,fig:ecdf4}) in one-sample Kolmogorov-Smirnov testing against the uniform distribution on $[0,2\pi]$.

\section{Conclusions}

The zero-noise limit of a stochastic perturbation of the point vortex system was investigated numerically in the case of configurations of three vortices leading to self-similar collapse at finite time under the deterministic dynamics. Compared to deterministic regularization obtained by smoothing the driving vector field, which selects a single continuation after collapse consisting in a self-similar burst of the same three vortices out of the collapse point,
the zero-noise limit produces a probability distribution supported by bursting self-similar trajectories preserving first integrals of the motion determined by the initial configuration before collapse.

When two symmetry classes are possible for self-similar evolution after collapse, the zero-noise limit selects one of them with equal probability. Moreover, conditionally to such similarity class, the (unique up to rotation) bursting solution after collapse time is rotated by a random angle, whose distribution is not uniform on $[0,2\pi]$, and appears to be sensible to the specific initial configuration of PV dynamics.

\begin{acknowledgments}
 The authors are indebted to many colleagues that contributed with their insight to the discussion on vortex dynamics after collapse, in particular T. Drivas, M. Maurelli, F. Flandoli, U. Pappalettera.
\end{acknowledgments}

\bibliography{biblio}

\end{document}